\documentstyle[twocolumn,aps,psfig]{revtex}

\begin{document}
\draft

\twocolumn[\hsize\textwidth\columnwidth\hsize\csname
@twocolumnfalse\endcsname

\title{The Extension of Rod-Coil Multiblock Copolymers\\  
and the Effect of the Helix-Coil Transition}
\author{A. Buhot and A. Halperin}

\address{UMR 5819 (CEA, CNRS, UJF), DRFMC/SI3M, 
CEA-Grenoble, 17 rue des Martyrs,\\
38054 Grenoble Cedex 9, France}
\date{\today }
\maketitle

\begin{abstract}
The extension elasticity of rod-coil mutliblock 
copolymers is analyzed for two experimentally 
accessible situations. In the quenched case, when 
the architecture is fixed by the synthesis, the 
force law is distinguished by a sharp change 
in the slope. In the annealed case, where 
interconversion between rod and coil states 
is possible, the resulting force law is sigmoid
with a pronounced plateau. This last case is 
realized, for example, when homopolypeptides 
capable of undergoing a helix-coil transition 
are extended from a coil state. 
Both scenarios are relevant to the analysis 
and design of experiments involving single 
molecule mechanical measurements of biopolymers
and synthetic macromolecules.
\end{abstract}


]

\narrowtext

With the advent of single molecule mechanical 
measurements it became possible to study the 
force laws characterizing the extension of 
individual macromolecules~\cite{rev}. 
In turn, these provide a probe of internal degrees
of freedom associated with intrachain self assembly 
or with monomers that can assume different 
conformational states. A molecular interpretation of
the force laws thus obtained requires appropriate 
theoretical models allowing for the distinctive 
``internal states'' of each system. The formulation 
of such models is a challenging task in view of the 
complexity and diversity of the systems investigated. 
These include DNA~\cite{DNA}, the muscle protein 
titin~\cite{titin} and the extracellular matrix 
protein tenascin~\cite{tenascin}, the polyscharides 
dextran~\cite{dextran} and xanthan~\cite{xanthan} 
as well as the synthetic polymer 
poly(ethelene-glycol)~\cite{PEG}. 

In this letter we consider two unexplored yet 
accessible systems where a detailed confrontation 
between theory and experiment is possible. In
particular, we present a theory for the extension 
force law of multiblock copolymers consisting of 
alternating rod and coil blocks. Two scenarios are
considered, both focusing on the equilibrium force 
law of long chains undergoing quasistatic extension. 
In one, the architecture is ``quenched'' that is, 
the block structure is set by the chemistry and no 
interconversion is possible. Such is the case, for 
example, for segmented polyurethans~\cite{Eisele}. 
In the second scenario the monomers can interconvert 
between coil and rod states. The ``annealed'' 
architecture is realized, for instance, in
homopolypeptides capable of undergoing a cooperative 
helix-coil transition~\cite{BP,PS,GK}. 
In this system, the highly rigid helical
domains play the part of\ the rod blocks. 
As we shall see, the two scenarios lead to 
distinctive force laws (Figure). The force law of 
the quenched case is characterized by an abrupt 
change of slope. This arises because the rod
blocks are more susceptible to orientation by the 
applied tension. In the annealed scenario, the 
extension of a chain that is initially in a coil
state leads to a sigmoid force profile exhibiting 
a pronounced plateau. The plateau is traceable to 
a one dimensional coexistence of helical and coil
domains where the chain extension favors the helical 
state because of its low configurational entropy. 
These results suggest that force measurements
can be used to probe the block structure of polymers 
with quenched architecture. In the annealed case 
the force law provides a direct measure
of the thermodynamic parameters involved. 
From the view point of the growing field of single 
molecule mechanical measurements, these results 
are helpful in exploring the diagnostic potential 
of these techniques. The discussion is of interest 
from the polymer physics perspective because the 
theory of polymer elasticity focuses on the case 
of flexible homopolymers modeled as random walks, 
or self avoiding random walks, with a constant step 
length and no internal degrees of freedom~\cite{GK}. 
In this context, rod-coil multiblock copolymers with 
quenched architecture may be viewed as a special
case of heteropolymers incorporating monomers of 
different sizes. The analysis of the annealed case 
supplements the isolated discussions of the
effect of internal degrees of freedom on the extension 
elasticity~\cite{Hill,Cluzel,BH,Gaub}.

We first consider the elastic free energy $F_{el}$ 
and the tension $f =\partial F_{el}/\partial R$ of 
a quenched rod-coil multiblock copolymer
with an imposed end-to-end distance, $R$. 
The chain incorporates $N$ monomers of identical 
length $a$ that form $yN$ rod blocks and $yN$ coil
blocks such that the number of monomers in the rods 
is $\theta N$. For simplicity we assume that the 
coil and rod blocks are monodispersed, and
that the number of monomers in a rod block is 
thus $\theta /y$. We focus on the case of long 
rod blocks, $\theta /y \gg 1$. The distinctive 
feature of rod-coil multiblock copolymers is that 
the rodlike blocks, the longer ``monomers'', are 
oriented by the applied tension before the alignment 
of the shorter monomers becomes significant. 
The freely jointed chain model, the macromolecular 
analog of the Langevin's theory of paramagnetism, 
enables a quantitative description of this 
situation~\cite{Hill,foot1}. The length of the 
monomer is the counterpart of the magnetic 
moment, the applied tension is the analog of 
the magnetic field while the end-to-end distance 
of the chain corresponds to the magnetization. 
A monomer of length $l$ is assigned an orientational 
energy of $-fl \cos \phi $ where $\phi $ is the
angle made by ${\bf l}$ with respect to ${\bf f}$. 
For a flexible homopolymer the end-to-end distance 
is $R=NaL(x)$ where $L(x)=\coth x-x^{-1}$ is the
Langevin function and $x=fa/kT$. In our situation 
the multiblock copolymer is viewed as a collection 
of two types of non-interacting ``dipoles'': 
$(1-\theta) N$ monomers of length $a$ and $yN$ 
``monomers'' of length $\theta a/y$. 
The reduced end-to-end distance, $r=R/Na$ is thus 

\begin{figure}
\centerline{\psfig{figure=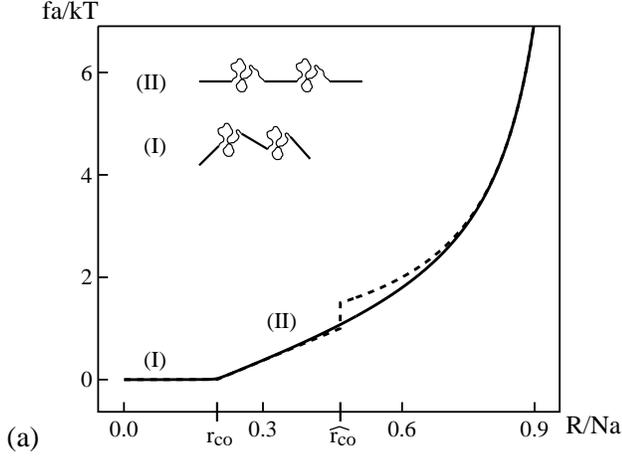,width=7.5cm}}
\centerline{\psfig{figure=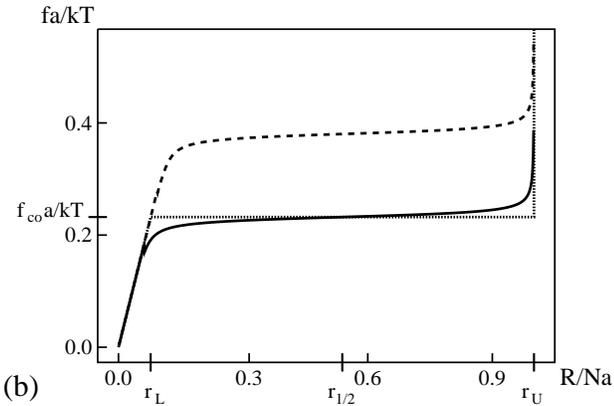,width=8cm}}
\caption{The force laws characterizing the extension 
of a rod-coil multiblock copolymer with a quenched 
(a) and an annealed (b) architectures. 
The parameters for (a) are $\theta = 0.2$ and 
$y = 10^{-4}$. The continuous line depicts the 
exact force law while the dashed line describes the 
approximate expression discussed in the text. 
The chain configurations in regions 
(I) and (II) are schematically depicted above. 
The parameters for (b) are $s = 0.7$ (dashed line) 
and $s = 0.8$ (continuous line). In both cases 
$\sigma = 10^{-4}$. The dotted line depicts the 
$S_{mix} = 0$ approximation for $s = 0.8$.
$f_{co}$, $r_L$, $r_{1/2}$ and $r_U$ correspond 
to $s=0.8$ case.}
\end{figure}

\begin{equation}
r=(1-\theta )L(x)+\theta L(\theta x/y).  \label{1}
\end{equation}

\noindent Three regimes are involved. 
When $x \ll 1$ and $\theta x/y \ll 1$ both the 
rods and the $a$ monomers are weakly oriented. 
Since $L(z) \approx z/3$ for $z \ll 1$, we find 
$fa/kT \approx 3r/(1-\theta +\theta^{2}/y)$ and 
$F_{el}/NkT \approx 3r^{2}/2(1-\theta +
\theta^{2}/y)$. For stronger extensions 
$x \ll 1$ while $\theta x/y \gg 1$ in other 
words, the $a$ monomers are only weakly aligned 
but the rods are almost fully oriented with $f$. 
Focusing on the case of long rods, $y \ll 1$, 
when $L(\theta x/y) \approx 1$, we obtain $fa/kT
\approx 3(r-\theta )/(1-\theta )$ and $F_{el}/NkT 
\approx 3(r-\theta )^{2}/2(1-\theta )+3y/2$. 
The first term in $F_{el}$ reflects 
the Gaussian stretching penalty of the 
$(1-\theta )N$ ``short'' monomers. 
It allows for the full alignment of the rods and 
the resulting modification of the imposed 
end-to-end distance experienced by the flexible
blocks. The second term reflects the Gaussian 
penalty associated with the full alignment of 
the $yN$ rod blocks. The crossover between 
the two regimes occurs at $r_{co} \approx 
\theta +y(1-\theta )/\theta $. Eventually, upon
further extension, the $a$ monomers also approach 
saturation, $x\gg 1$ and $\theta x/y \gg 1$. In this 
last regime $fa/kT \approx (1-\theta +y)/(1-r)$ and 
$F_{el}/NkT\approx (1-\theta +y)\ln [2(1-\theta )/3(1-r)]
+3y/2+(1-\theta )/6$. Since the approximate expressions 
for $f$ do not match at the boundary between the two 
last regimes, the crossover between them, 
$\widehat{r_{co}}$, may be determined by the requirement 
that $fa/kT \approx 3(r-\theta)/(1-\theta) \approx 1$ 
leading to $\widehat{r_{co}} \approx (1+2\theta )/3$. 
Altogether, the plot of $fa/kT$ vs. $r$ for a rod-coil 
multiblock copolymer is distinguished by an abrupt 
change of slope at $r_{co}$. In turn, this feature 
provides a useful diagnostic for the architecture of 
the polymer.

In the annealed scenario, monomers may interconvert 
between the rod and coil states. 
Accordingly, $\theta$ and $y$ are no longer constants 
set by the chemistry. Rather, their values vary with 
the temperature $T$ and with $R$. $\theta$ and $y$ 
are determined by minimizing the free energy per 
chain $F_{chain}$ with respect to $y$ and $\theta$ 
for a given $R$. $F_{chain}$ allows, in addition to 
$F_{el}$, for the mixing free energy $F_{mix}$ of
the one dimensional mixture, along the chain 
trajectory, of the rod and coil blocks. 
Our discussion focuses on polymers undergoing a 
helix-coil transition. It is directed at the 
coupling of the helix-coil transition with
the extension for the simplest possible situation, 
that is a homopolypeptide capable of forming an 
$\alpha$-helix~\cite{BP,PS,GK}. This form
of intrachain self assembly is due to intrachain 
hydrogen bonds. The $i$-th monomer, residue, 
forms H-bonds with the $(i+3)$ and $(i-3)$ monomers.
Altogether, a helical domain consisting of $n$ 
monomers involves $n-2$ H-bonds. 
The persistence length of an $\alpha$-helix is of 
order of $200 nm$ and a helical domain may thus 
be considered as a rod block. It is convenient
to discuss this system in terms of ``bonds'' 
between two adjacent monomers. For simplicity we 
retain the approximation setting the monomer size in 
the rod and coil state equal to $a$. Using the coil 
state as a reference, each helical bond is assigned 
a free energy $\Delta f$. It reflects the contribution 
of the intrachain H-bonds, the change in solvation due 
to the formation of intrachain H-bonds and the 
associated loss of configurational entropy. 
The helix-coil transition occurs at the transition 
temperature $T_{\ast}$ when $\Delta f=0$. 
Above $T_{\ast}$, $\Delta f>0$ while below $T_{\ast}$, 
$\Delta f<0$. Terminal helical bonds, at the boundary 
of a helical domain, incur an additional free energy 
penalty $\Delta f_{t}>0$ since the terminal monomers 
lose their configurational entropy but do not contribute 
H-bonds. $\Delta f_{t}$ plays the role of interfacial 
free energy associated with the helix-coil domain boundary. 
Traditionally, the theory of the helix-coil transition is 
formulated in terms of the Bragg-Zimm parameters $s = \exp 
(-\Delta f/kT)$ and $\sigma = \exp (-2 \Delta f_{t}/kT)$
where $\sigma \approx 10^{-3}-10^{-4}$ varies with the 
identity of the residues but is independent of $T$. 
The helix-coil transition gives rise to a sigmoid 
$\theta$ vs. $T$ plot with a characteristic width 
$\sim T_{\ast} \sigma^{1/2}$. The usual discussion 
of the helix-coil transition is based on transfer 
matrix methods. For our purposes it is convenient to 
recast it in terms of the free energy of the unperturbed 
chain, $F_{chain}^{0}$. For long chain, when the $N 
\rightarrow \infty$ limit is applicable, $F_{chain}^{0}$ 
reflects three contributions. Each of the $\theta N$ 
helical bonds contributes $\Delta f$ while every one 
of the $2yN$ terminal bonds contributes $\Delta f_{t}$. 
These two terms are supplemented by the mixing
entropy, $S_{mix}$, associated with the different 
possible combinations of $\theta N$ helical bonds and 
$2yN$ terminal bonds. The number of possible configurations, 
$W$ is $W_{th}W_{tc}$ where $W_{th} = {\theta N \choose yN}$
is the number of ways of placing $yN$ terminal bonds 
among $\theta N$ helical bonds and $W_{tc} = {(1-\theta) 
N \choose yN}$ is the statistical weight associated with 
the placement of $yN$ terminal bonds among $(1-\theta)N$ 
coil bonds. $S_{mix}=k \ln W$ can be expressed as $S_{mix}  
= \theta S_{mix}(y/\theta)+(1-\theta )S_{mix}(y/(1-\theta))$ 
where $\frac{1}{N} S_{mix}(X) = - X \ln X - (1-X) \ln 
(1-X)$. Here $\frac{1}{N} S_{mix}(y/\theta)$ is the mixing 
entropy of the terminal monomers among the rod monomers while 
$\frac{1}{N} S_{mix}(y/(1-\theta))$ is the mixing entropy of the 
terminal monomers among the coil monomers. Altogether 

\begin{eqnarray}
\frac{F_{chain}^{o}}{NkT} & = & \theta \frac{\Delta f}{kT}
+ 2y \frac{\Delta f_{t}}{kT} + (\theta-y) \ln 
\frac{\theta -y}{\theta}+y \ln \frac{y}{\theta} 
\label{2} \\
& & +(1-\theta -y) \ln \frac{1-\theta -y}{1-\theta}
+y \ln \frac{y}{1-\theta}. \nonumber
\end{eqnarray}

\noindent The equilibrium conditions $\partial 
F_{chain}^{0}/\partial y =0$ and $\partial 
F_{chain}^{0}/\partial \theta = 0$ lead respectively 
to $y^{2}=(\theta -y)(1-\theta -y) \sigma$ and to 
$(1-\theta )(\theta -y) = \theta (1-\theta -y) s$.
In turn, these yield the familiar results for the 
unperturbed chain and, in particular, $\theta = 
\frac{1}{2} + \frac{1}{2} \sqrt{\frac{(s-1)^{2}}{4 
\sigma s+(s-1)^{2}}}$.

The full analysis of the coupling of the helix-coil 
transition with the extension of the chain involves 
an augmented free energy per chain, allowing for the 
extension penalty $F_{chain} = F_{chain}^{0} + 
F_{el}$~\cite{foot2}. An analytic solution yielding 
the qualitative features of the force law, as well
as the important length and force scales, is possible 
by setting $S_{mix}=0$~\cite{BH}. Within the $S_{mix}=0$ 
approximation the interfacial penalty imposes coalescence 
of the helical domains leading to the formation of a
helix-coil diblock copolymer ($y=0$). As a result, the helix-coil 
transition takes place as a first order phase transition. 
The corresponding free energy is 

\begin{equation}
\frac{F_{chain}}{NkT} \approx \frac{3(r-\theta)^2}{2
(1-\theta)} + \theta \frac{\Delta f}{kT}.  \label{3}
\end{equation}

\noindent The equilibrium condition $\partial 
F_{chain}/\partial \theta =0$ for a given $r$ yields 
$\theta = 1 - (1-r)/\sqrt{1-2\Delta f/3kT}$. 
In turn, this defines a critical extension, $r_{L} 
= 1 - \sqrt{1-2\Delta f/3kT}$ such that for $r<r_{L}$,  
$\theta = 0$ and for $r \geq r_{L}$, $\theta$ is 
specified by $\theta = (r-r_{L})/(1-r_{L})$. 
The equilibrium free energy of the chain for $r<r_{L}$ 
is $F_{chain}/NkT \approx 3 r^2/2$ while for $r \geq r_L$
it is $F_{chain}/NkT \approx 3r_{L}(r-1)+\Delta f/kT$. 
This regime lasts until $\theta = 1$ is attained at 
$r_{U}=1$. Within this simple minded picture, further 
extension is impossible and for $r_{U}>1$, $F_{chain}
= \infty$. The corresponding force law involves three 
regimes: (i) A linear response regime, where the Gaussian 
elasticity is operative and $fa/NkT \approx 3 r$, occurs 
while $r<r_{L}$ and $\theta = 0$. (ii) A plateau with 
$f_{co}a/NkT \approx 3 r_{L}$ occurs in the range 
$r_{L}<r<r_{U}$ when the coexistence between the helix 
and coil states follows the lever rule $\theta/(1-\theta)
= (r-r_{L})/(r_{U}-r)$. The midpoint of the plateau,
corresponding to $\theta = 1/2$, occurs at $r_{1/2}
= (r_U+r_L)/2$. (iii) A steep increase in force occurs 
at $r = r_{U}$, when the assumed nonextensibility of 
the fully helical chain comes into play. Note that in
reality the helical domains are not perfect rods. 
High applied tension may cause extension by breaking 
the H-bonds. Within the $S_{mix}=0$ approximation 
the plateau corresponds to a first order phase transition
involving the coexistence of a helical phase and a coil phase. 
Such phase transition is prohibited in one dimensional 
system experiencing short ranged interactions~\cite{LL}. 
When one allows for $S_{mix}>0$ the plateau in the
force law exhibits a weak increase with $r$, instead 
of the $r^{0}$ dependence predicted by the $S_{mix}=0$ 
approximation. Yet, as we shall see, the center of the 
plateau occurs at $r_{1/2}$ and its height at this point
is $f_{co}a/kT \approx 3 r_L$. Furthermore, the qualitative 
$\Delta f$ dependence of both the height and the width of 
the plateau is retained. In particular, the width of the 
plateau increases while its height decreases as $\Delta 
f \rightarrow 0$, or equivalently, as $s \rightarrow 1$. 
Since at the vicinity of the transition temperature 
$s \sim (T-T_{\ast})/T_{\ast}$, these scenarios may be 
explored by change of $T$. 

The complete analysis of the problem in terms $F_{chain}
= F_{chain}^{0}+F_{el}$ involves different regimes 
distinguished by the applicable form of $F_{el}$,
as discussed in the first part of this letter. 
For weak extensions $F_{el}/NkT \approx 3r^{2}/2 
(1-\theta +\theta ^{2}/y)$ and the equilibrium
conditions $\partial F_{chain}/\partial \theta = 0$ 
and $\partial F_{chain}/\partial y = 0$ lead to 
$\partial F_{chain}^{0}/\partial \theta = - \partial 
F_{el}/\partial \theta$ and $\partial F_{chain}^{0}/\partial
y = - \partial F_{el}/\partial y$. In turn, this leads to 
$y^{2}=(\theta-y)(1-\theta-y) \sigma K_{\sigma}$ and to 
$(1-\theta)(\theta-y)=\theta (1-\theta -y) s K_{s}$ where 
$K_{\sigma} = \exp [(\partial F_{el}/\partial y)/NkT]$ and 
$K_{s} = \exp [(\partial F_{el}/\partial \theta )/NkT]$. 
$(\partial F_{el}/\partial \theta)/NkT$ is the increment in 
the elastic free energy per monomer associated with adding one 
residue to a helical block for a constant $y$ and $(\partial 
F_{el}/\partial y)/NkT$ is the price of creating an extra 
helical sequence while maintaining a constant $\theta$.
For weak extensions, when the elastic penalty is a weak 
perturbation, $K_{\sigma}$ and $K_{s}$ may be approximated 
by $K_{\sigma} = \exp [-r^{2} \theta_{0}^{2}/y_{0}^{2} 
(1-\theta _{0}+\theta _{0}^{2}/y_{0})^{2}]$
and $K_{s} = \exp [r^{2}(2\theta_{0}/y_{0}-1)/(1-\theta_{0}
+\theta_{0}^{2}/y_{0})^{2}]$ where $\theta_{0}$ and $y_{0}$ 
correspond to the unperturbed chain. $K_{\sigma}$ is a 
decreasing function of $r$ while $K_{s}$ is an increasing 
function of $r$. Thus, the extension favors the transition
by increasing $s$ and enhances the cooperatively by decreasing 
$\sigma$. This analysis demonstrates that the $S_{mix}=0$ 
approximation overestimates $F_{el}$ and thus $f$. 
$fa/NkT \approx 3r/(1-\theta _{0}+\theta _{0}^{2}/y_{0})$
provides a better estimate for $f$. However, in accordance 
with the Le Chatellier principle~\cite{LL}, the actual force 
law is even weaker since $\theta$ increases with $r$ while 
$y/\theta$ decreases. For stronger deformations, when 
$r>r_{co}$ the appropriate elastic free energy is 
$F_{el}/NkT \approx 3(r-\theta )^{2}/2(1-\theta )+3y/2$. 
In this range it is possible to obtain the force law 
at the vicinity of $\theta =1/2$. The equilibrium condition 
$\partial F_{chain}/\partial y = 0$ leads to $y^{2} = 
(\theta -y)(1-\theta -y) \sigma/e^{3/2}$. Since $\sigma \ll 1$ 
this reduces, in the vicinity of $\theta =1/2$, 
to $y \approx \sqrt{\theta (1-\theta) \sigma/e^{3/2}}$. 
Utilizing this relationship, and following expansion of the
logarithmic terms we obtain $F_{chain}/kT \approx 
3(r-\theta )^{2}/2(1-\theta) +\theta \Delta f/kT 
- 2\sqrt{\theta (1-\theta) \sigma/e^{3/2}}$. 
In turn, $\partial F_{chain}/\partial \theta = 0$ recovers 
the expression for $\theta$ as obtained within the $S_{mix}=0$ 
approximation, $\theta \approx (r-r_L)/(1-r_L)$. 
This allows us to express $F_{chain}$ as a function of $r$ 
and to obtain $f$ in the vicinity of $\theta = 1/2$ 

\begin{equation}
\frac{fa}{kT} \approx \frac{f_{co}a}{kT} + 2 \  
\frac{\sigma^{1/2}}{e^{3/4}} \frac{(r-r_{1/2})
(1-r_L)^{-1}}{\sqrt{(r-r_L)(r_U-r)}}. \label{4}
\end{equation}

\noindent The force at $r_{1/2}$ is $f_{co}$ as suggested 
by the $S_{mix}=0$ approximation. However, when the full 
$F_{chain}$ is allowed for the $f \sim r^{0}$ behavior of 
a perfect plateau is lost. The second term in (\ref{4}) 
ensures that $f$ increases with $r$. 
On the other hand, since $\sigma \approx 10^{-4}$, 
the $\sigma^{1/2}$ prefactor in this term imposes slow 
variation of $f$ in the vicinity of $r_{1/2}$.

Our discussion of the elasticity of rod-coil multiblock 
copolymers reveals two scenarios. 
In the quenched case, the force law exhibits a sharp change
of slope while in the annealed case the force law is sigmoid. 
It is important to note that our analysis focused on the 
thermodynamic force law as obtained for long chains and 
quasistatic extension. Effects due to the finite length of 
the chains and the finite rate of deformation may modify
the experimentally observed force curves~\cite{Gaub}. 
The analysis of the coupling between the helix-coil 
transition and the chain extension focused on the case of 
homopolypeptides capable of forming a single strand 
$\alpha$-helix. The results can be however applied, 
with some modifications, to a much wider class of systems. 
It is relevant, for example, to the interpretation of the 
sigmoid force curve observed for poly(ethylene-glycol)
which was attributed to the formation of water-mediated 
single stranded helix~\cite{PEG}. 
Furthermore, the analysis can be extended to the case of
polymers forming multistranded helices such as collagen. 
In this context it is of interest to note that sigmoid 
force curves were indeed observed in early experiments 
involving fibers formed by fibrous proteins, collagen,
keratin etc., that form helical structures~\cite{Flory}. 
Finally, one should note the recent development of facile 
synthesis route for homopolypeptides and block 
copolypeptides of well defined architecture~\cite{TD}. 
This provides a convenient method for synthesizing rod-coil 
multiblock copolymers with either quenched or annealed 
architecture.


\begin{references}
\bibitem{foot1}  
By adopting the freely jointed chain model we ignore
possible effects due to the self avoidance of the chain, 
that is the Pincus elasticity.
\bibitem{foot2}  
Within this model, the derivation of $F_{el}$ and 
$S_{mix}$ involve different assumptions. $F_{el}$ 
is obtained for monodispersed blocks while $S_{mix}$ 
reflects the polydispersity of the domains.
\end{references}
\end{document}